\documentclass[aps,prl,reprint,groupedaddress,amsmath]{revtex4-1}
\usepackage{graphicx}
\usepackage{amssymb}

\begin{document}
\title{The spin-$\frac{1}{2}$ XXZ chain system Cs$_2$CoCl$_4$ in a transverse magnetic field
}
\author{O. Breunig$^1$}

\author{M. Garst$^2$}
\author{E. Sela$^{2,3}$}
\author{B. Buldmann$^2$}

\author{P. Becker$^4$}
\author{L. Bohat\'y$^4$}
\author{R. M\"uller$^1$}

\author{T. Lorenz$^1$}

\affiliation{$^{1}$II. Physikalisches Institut, Universit\"at zu K\"oln, Z\"ulpicher Str. 77, 50937 K\"oln, Germany}
\affiliation{$^{2}$Institut f\"ur Theoretische Physik, Universit\"at zu K\"oln, Z\"ulpicher Str. 77, 50937 K\"oln, Germany}
\affiliation{$^{3}$Raymond and Beverly Sackler School of Physics and Astronomy, Tel-Aviv University, Tel Aviv 69978, Israel}
\affiliation{$^{4}$Institut f\"ur Kristallographie, Universit\"at zu K\"oln, Greinstra\ss{}e 6, 50939 K\"oln, Germany}

\begin{abstract}
Comparing high-resolution specific heat and thermal expansion measurements to exact finite-size diagonalization, we demonstrate that Cs$_2$CoCl$_4$ for a magnetic field along the crystallographic $b$ axis realizes the spin-$\frac{1}{2}$ XXZ chain in a transverse field. Exploiting both thermal as well as virtual excitations of higher crystal field states, we find that the spin chain is in the XY-limit with an anisotropy $J_z/J_\perp \approx 0.12$ substantially smaller than previously believed. A spin-flop Ising quantum phase transition occurs at a critical field of $\mu_0 H_b^{\rm cr} \approx 2$ T before around $3.5$ T the description in terms of an effective spin-$\frac{1}{2}$ chain becomes inapplicable.
\end{abstract}
\maketitle


Low-dimensional spin-systems in the presence of an applied magnetic field provide paradigmatic examples of quantum phase transitions. Such transitions occur whenever the magnetic field induces a change in the magnetic ground state which is then reflected at finite temperatures in metamagnetic behavior and anomalous thermodynamics. A prominent representative is the quantum phase transition in the Ising universality class that can be explicitly realized with effective Ising-spin chains in a transverse magnetic field as, e.g., in the celebrated LiHoF$_4$ or CoNb$_2$O$_6$ \cite{Bitko1996,Coldea2010}. 
The Ising transition, however, also emerges in XY-spin chains in the presence of a transverse field, i.e., a magnetic field within the XY-plane. At small transverse fields,  a spin-flop phase spontaneously breaks the remaining Ising symmetry that is then restored at larger fields when entering the field-polarized phase.

\begin{figure}[t]
 \centering
\includegraphics[width=.9\columnwidth]{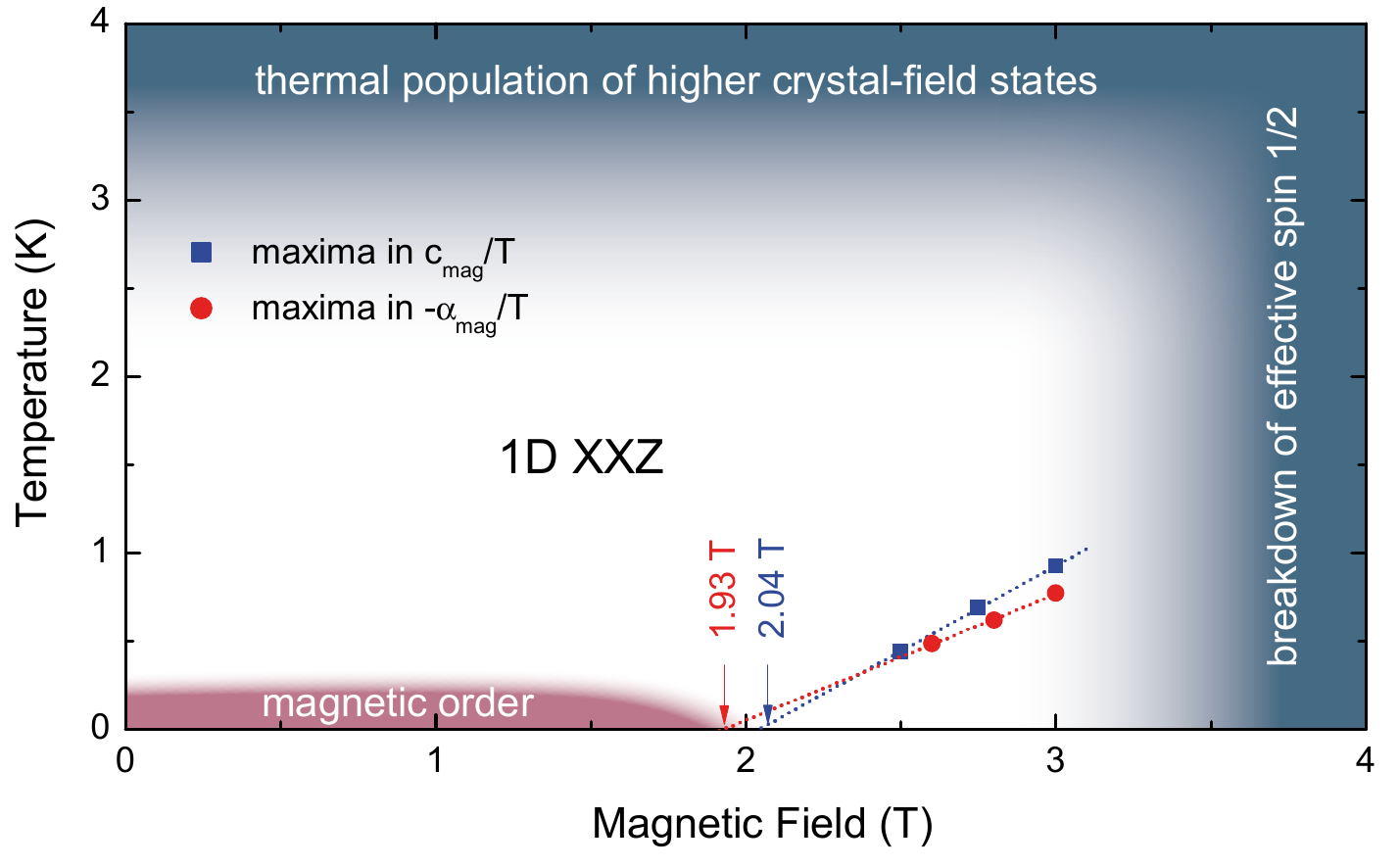}
\caption{(color online) Phase diagram of Cs$_2$CoCl$_4$ for a magnetic field along the \textit{b} axis. An extended regime (1D XXZ) is described by the spin-$\frac{1}{2}$ XXZ chain Hamiltonian in transverse field \eqref{SpinChain}. Crossover lines are characteristic for Ising criticality $T \sim |H_b -H^{\rm cr}_b|^{\nu z}$ with $\nu z=1$ and extrapolate to a critical field  $\mu_0 H^{\rm cr}_b \approx 2$ T. 
 }
 \label{fig1}
\end{figure}

An Ising transition of the latter type is approximately realized in the spin-$\frac{1}{2}$ XXZ chain system Cs$_2$CoCl$_4$ \cite{Algra1976,McElearney1977,smit1979field,duxbury1981transverse, Kenzelmann2002,Chatterjee2003} at a moderate critical  field $\mu_0 H_b^{\rm cr} \approx 2$ T, see Fig.~\ref{fig1}. 
The magnetism in Cs$_2$CoCl$_4$ arises from the Co$^{2+}$ ions that according to Hund's first rule realize an $S=3/2$ groundstate. Each Co$^{2+}$ is embedded in a distorted tetrahedron of four chlorine atoms so that the corresponding crystal field results in a single-ion anisotropy $D$. The CoCl$_4$ tetrahedra form chains with dominant Heisenberg exchange $J_\mathrm{H}$ along the crystallographic $b$ axis. Neglecting interchain interactions, the magnetism is thus described in terms of a spin-$\frac{3}{2}$ chain,
\begin{align} \label{Ham1}
\mathcal{H}_{3/2} = \sum_{i} \Big( J_\mathrm{H} \vec{\mathcal{S}}_i  \vec{\mathcal{S}}_{i+1} + D (\mathcal{S}^z_i)^2  - g_{\frac{3}{2}} \mu_0 \mu_\mathrm{B} H_b \mathcal{S}^x_i \Big),
\end{align}
where $\vec{\mathcal{S}}$ is a spin-$\frac{3}{2}$ operator. For the coupling constants, we find 
\begin{align} \label{parameters}
J_\mathrm{H}/k_\mathrm{B} = 0.74\, {\rm K},\; D/k_\mathrm{B} = 7.0\, {\rm K}, \mbox{ and } g_{\frac{3}{2}} = 1.94.
\end{align} 
In Cs$_2$CoCl$_4$, a complication arises from an alternating orientation of the CoCl$_4$ tetrahedra in neighboring chains along the $c$ direction. This results in two types of easy planes within a single unit cell and identifies the $b$ axis as the only common in-plane direction for all sites. Only a magnetic field along this $b$ axis translates to a non-staggered in-plane field $H_b$.

The strong single-ion anisotropy, $D \gg J_\mathrm{H}$, splits the $S=3/2$ quartet of a single ion into two Kramers doublets at $H_b=0$. The thermal population of the higher-energy doublet $|\pm\frac{3}{2}\rangle$ gives rise to a Schottky anomaly, i.e., a contribution to the 
molar specific heat of the form
\begin{align} \label{Schottky}
c_{\rm Sch}(T) = N_{\rm A} k_\mathrm{B} \left(\frac{\Delta E}{k_\mathrm{B} T}\right)^2 
\frac{e^{- \Delta E/(k_\mathrm{B} T)} }{(1+e^{- \Delta E/(k_\mathrm{B} T)})^2}.
\end{align}
In zeroth order in $J_\mathrm{H}$ and small magnetic fields $g_{\frac{3}{2}} \mu_0 \mu_B H_b \ll D$, the energy gap is  given by the anisotropy, 
$\Delta E = 2 D$.

At low temperatures $k_\mathrm{B} T \ll \Delta E$, when the higher-energy doublet is frozen out, a description of the low-energy doublet  $|\pm\frac{1}{2}\rangle$ in terms of an effective spin-$\frac{1}{2}$ XXZ chain arises,
\begin{align}
\label{SpinChain}
\lefteqn{\mathcal{H}_{\rm XXZ} =}
\\ \nonumber
&\sum_i \Big[ J_\perp (S_i^x S_{i+1}^x+S_i^y S_{i+1}^y ) + J_z S_i^z S_{i+1}^z - g \mu_0 \mu_\mathrm{B} H_b S_i^x \Big].
\end{align}
Performing a Schrieffer-Wolff transformation on the Hamiltonian \eqref{Ham1} up to first order in $1/D$ we obtain in the limit of small fields $g_{\frac{3}{2}} \mu_0 \mu_\mathrm{B} H_b \ll D$ the relations
\begin{align} \label{constants}
J_\perp = 4 J_\mathrm{H},\; J_z = J_\mathrm{H} - \frac{39}{8} \frac{J_\mathrm{H}^2}{D} ,\;
g = 2 g_{\frac{3}{2}}\left(1 - \frac{3}{2} \frac{J_\mathrm{H}}{D} \right). 
\end{align}
Besides the thermal activation, which leads to the Schott\-ky anomaly Eq.~\eqref{Schottky}, there are also virtual excitations of the higher-energy doublet $|\pm\frac{3}{2}\rangle$ that renormalize $J_z$ and $g$ of the effective XXZ chain \eqref{SpinChain}. 
These corrections are negligible in the limit $J_\mathrm{H} /D \to 0$ where the resulting XXZ chain takes on an anisotropy $J_z/J_\perp \to \frac{1}{4}$. This limit was taken for granted in all previous experimental \cite{Algra1976,McElearney1977,smit1979field,duxbury1981transverse, Kenzelmann2002} as well as theoretical \cite{Caux2003,siahatgar2008thermodynamic} works on Cs$_2$CoCl$_4$. Although the ratio $J_\mathrm{H}/D \approx 0.1$ is small, we find, however, that it leads to a sizeable correction of $J_z$ due to the numerically large prefactor 39/8 in Eqs.~\eqref{constants} and reduces the anisotropy of the XXZ chain even further to $J_z/J_\perp \approx 0.12$.
We demonstrate in the following that this important correction arising from the virtual excitations of the higher-energy crystal field states is instrumental for a consistent description of thermodynamics.

In the limit of small fields the magnetic contribution to thermodynamics is well described by the spin-$\frac{1}{2}$ chain Hamiltonian \eqref{SpinChain} supplemented by the Schottky anomaly \eqref{Schottky}. Experimentally, we find that this description breaks down above a magnetic field of $\approx 3.5$~T, see Fig.~\ref{fig1}, where the $|\pm\frac{3}{2}\rangle$ and $|\pm\frac{1}{2}\rangle$ states start to become strongly entangled. In addition, at the lowest temperatures the weak interchain couplings become important and finally stabilize long-range antiferromagnetic order at $T_{\rm N} \approx 220$~mK \cite{Kenzelmann2002,Chatterjee2003}.

There exists thus an extended regime in temperature and field where the magnetism of Cs$_2$CoCl$_4$ is governed by the spin-$\frac{1}{2}$ XXZ chain Hamiltonian \eqref{SpinChain}. This Hamiltonian possesses two critical gapless points \cite{Kurmann1982,dmitriev2002one,Dmitriev2001}. For $H_b=0$ the ground state is a Luttinger liquid with algebraically decaying correlations. A finite field $H_b$ immediately quenches these correlations leading to a gap. The classical counterpart of the corresponding groundstate is the antiferromagnetic spin-flop  configuration \cite{Kurmann1982}.
Upon increasing the magnetic field further one encounters an Ising quantum phase transition into a field-polarized phase. This field-polarized phase is again gapped and characterized by quantum fluctuations which suppress the magnetization below the saturation limit of the effective spin $\frac{1}{2}$. Interestingly, the Ising quantum phase transition at $\mu_0 H^{\rm cr}_b \approx 2$~T is located within the validity regime of the spin-$\frac{1}{2}$ XXZ chain Hamiltonian. As a consequence, its Ising criticality prevails in Cs$_2$CoCl$_4$ close to $H^{\rm cr}_b$.

In the present work, we study the one-dimensional magnetism of Cs$_2$CoCl$_4$ experimentally by specific heat and thermal expansion measurements together with a comparison to the theoretical predictions of the XXZ chain.
Single crystals were grown from an aqueous solution of CoCl$_2\cdot$\,6H$_2$O and CsCl by slow evaporation. 
Specific heat was measured by a thermal relaxation method using a home-built calorimeter. Thermal expansion 
was measured along the crystallographic \textit{b} axis, i.e. parallel to the applied magnetic field, on a home-built ca\-pa\-ci\-tance dilatometer. 
For the quantitative comparison to theory, we performed exact diagonalization of the Hamiltonian \eqref{SpinChain} 
for finite chains with $N=18$ and $N=16$ sites at zero and finite field, respectively, using the ALPS code \cite{Bauer2011}. We carefully checked that our conclusions are independent of finite-size corrections. 

\begin{figure}
\centering
\includegraphics[width=1.\columnwidth]{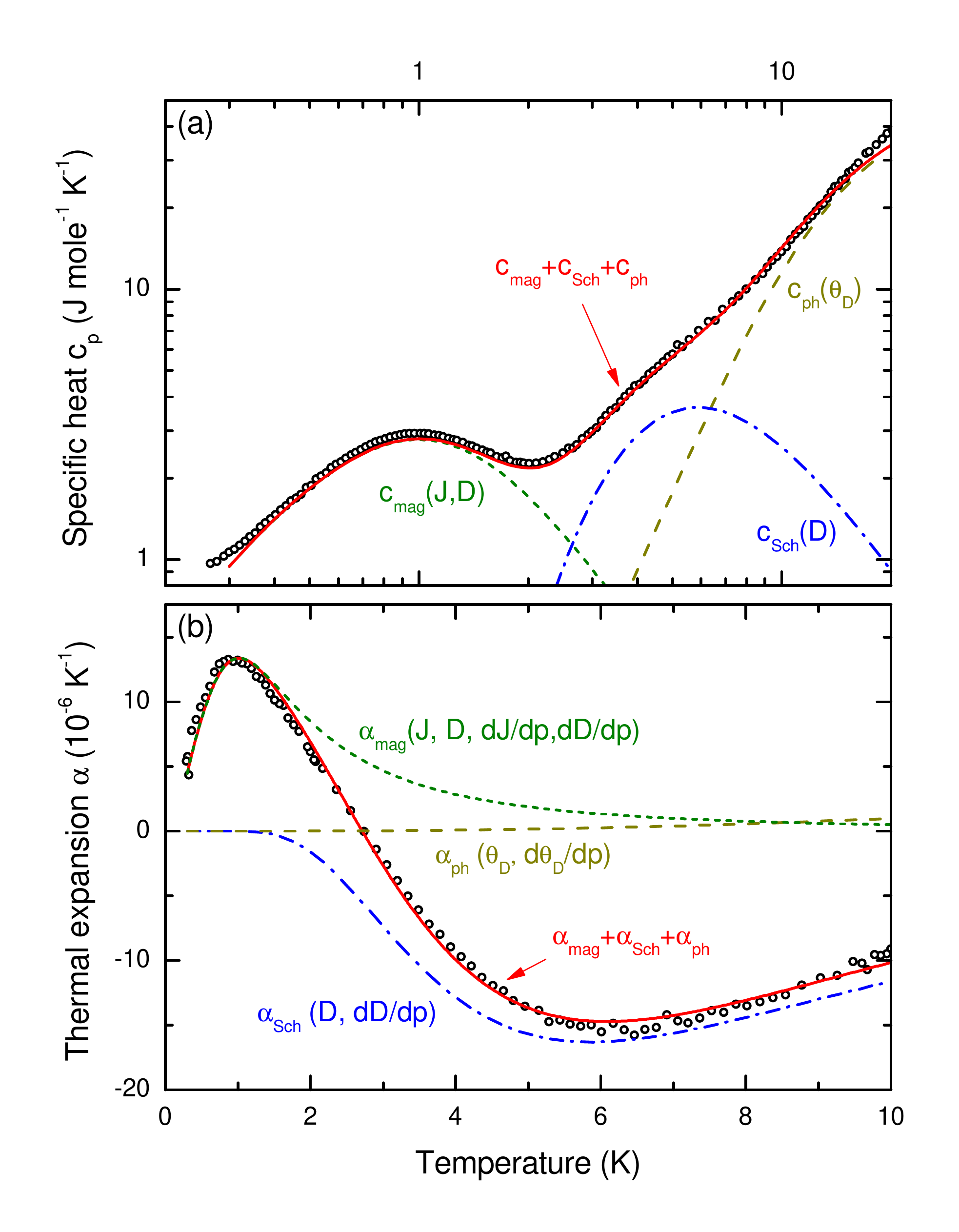}
\caption{(color online) Specific heat and \textit{b}-axis thermal expansion of Cs$_2$CoCl$_4$ in zero field (symbols). Both quantities can be decomposed into three contributions attributed to the phonons (long-dashed line), the Schottky anomaly (dashed-dotted line) and the XXZ chain (short-dashed line), see text.}
 \label{fig2}
\end{figure}

In the parameter regime of interest, the free energy density of Cs$_2$CoCl$_4$ can be approximated by three terms $\mathcal{F} = \mathcal{F}_{\rm ph} + \mathcal{F}_{\rm Sch} + \mathcal{F}_{\rm XXZ}$. The first contribution derives from the phonons and can be approximated by the Debye model. The latter two are attributed to the magnetic sector: $\mathcal{F}_{\rm Sch}$  describes the thermal occupation of the higher-energy crystal field states and the free energy density of the XXZ chain of $N$ spins is $\mathcal{F}_{\rm XXZ} = - k_\mathrm{B} T\ln$ tr$\{e^{-\mathcal{H}_{\rm XXZ}/(k_\mathrm{B} T)}\}/(N V_S)$ with the volume $V_S = 235$ \AA$^3$ per spin. 

The molar specific heat follows straightforwardly, $c = - V_m T \partial^2_T \mathcal{F}$ with the molar volume $V_m = N_{\rm A} V_S$, and, correspondingly, decomposes into three contributions  $c = c_{\rm ph} + c_{\rm Sch} + c_{\rm XXZ}$. The specific heat in zero field, see Fig.~\ref{fig2}(a), agrees with previously reported results within a few percent \cite{Algra1976}. 
The behavior at higher temperatures is governed by $c_{\rm ph}$. At lower temperatures the specific heat shows two features attributed to magnetism, i.e., the Schottky anomaly and the XXZ chain. 
Previous works have treated these two features independently
\cite{Algra1976,McElearney1977}.
However, it is important to realize that they derive from the same parent Hamiltonian \eqref{Ham1}, which is characterized by only two coupling constants, the Heisenberg coupling  $J_\mathrm{H}$ and the anisotropy $D$. 
The position of the Schottky anomaly essentially fixes the anisotropy $D$ and the low-energy peak in the specific heat then determines the remaining free parameter $J_\mathrm{H}$ via the relations \eqref{constants} for the coupling constants of the XXZ chain.
By consistently fitting the specific heat we obtain a best fit, shown as the red solid line in Fig.~\ref{fig2}(a), for a Debye temperature $\Theta_{\rm D} = 67$ K and the parameters $J_\mathrm{H}$ and $D$ given in Eqs.~\eqref{parameters}.
The values of the anisotropy $D$ and the corresponding $J_\perp=4J_\mathrm{H}$ agree well with those obtained in previous studies~\cite{Algra1976,McElearney1977,duxbury1981transverse}, but as mentioned above the anisotropy $J_z/J_\perp\approx 0.12$ is significantly smaller than assumed previously~\cite{Algra1976,McElearney1977,smit1979field,duxbury1981transverse, Kenzelmann2002,Caux2003,siahatgar2008thermodynamic}.
 
The interdependence of the Schottky-anomaly gap and the parameters of the XXZ chain becomes particularly significant when considering the thermal expansion. The thermal expansion along the $b$ axis is given by $\alpha =\partial^2 \mathcal{F}/(\partial p_b \partial T)$ where $p_b$ is uniaxial pressure along the $b$ axis.
In a perturbative treatment of the elastic coupling, the thermal expansion just derives from the pressure dependence of the coupling constants so that $\alpha$ also decomposes into three contributions $\alpha = \alpha_{\rm ph} + \alpha_{\rm Sch} + \alpha_{\rm XXZ}$, see Fig.~\ref{fig2}(b). 
As the phonon contribution and the Schottky anomaly only depend on single energy scales, they both obey Gr\"uneisen relations, $\alpha_{\rm ph}/c_{\rm ph} = \frac{1}{V_m} \partial \ln \Theta_{\rm D}/\partial p_b$ and $\alpha_{\rm Sch}/c_{\rm Sch} = \frac{1}{V_m} \partial \ln D/\partial p_b$. 
In contrast, the contribution from the XXZ chain depends on two coupling constants, $J_\mathrm{H}$ and $D$, 
and thus its Gr\"uneisen relation is, in principle, violated already in zero field, $\alpha_{\rm XXZ}/c_{\rm XXZ} \neq$ const..
In a finite magnetic field, $\alpha_{\rm XXZ}$ even comprises three terms
\begin{align} \label{alphaXXZ}
\lefteqn{\alpha_{\rm XXZ} =}\\\nonumber& 
\left( \frac{\partial^2 \mathcal{F}_{\rm XXZ}}{\partial T \partial J_\perp} \right) \frac{\partial J_\perp}{\partial p_b} 
+ \left( \frac{\partial^2 \mathcal{F}_{\rm XXZ}}{\partial T \partial J_z}\right) \frac{\partial J_z}{\partial p_b}
+ \left( \frac{\partial^2 \mathcal{F}_{\rm XXZ}}{\partial T \partial g}\right) \frac{\partial g}{\partial p_b}.
\end{align}
The prefactors of the first two terms can be identified with $T$-derivatives of equal-time bond correlators \cite{Anfuso2008,Zapf2008},
\begin{align}
\label{transCorr}
\frac{\partial^2 \mathcal{F}_{\rm XXZ}}{\partial T \partial J_\perp} 
&= \frac{1}{N V_S} \sum_{i}^N \frac{\partial}{\partial T} \langle   S_i^x S_{i+1}^x+ S_i^y S_{i+1}^y  \rangle,
\\
\label{longCorr}
\frac{\partial^2 \mathcal{F}_{\rm XXZ}}{\partial T \partial J_z}&=\frac{1}{N V_S} \sum_{i}^N \frac{\partial}{\partial T} \langle   S_i^z S_{i+1}^z \rangle,
\\
\frac{\partial^2 \mathcal{F}_{\rm XXZ}}{\partial T \partial g}&=-\frac{\mu_0 \mu_\mathrm{B} H_b}{N V_S} \sum_{i}^N \frac{\partial}{\partial T} \langle   S_i^x  \rangle.
\end{align}
The prefactor of the last term in Eq.~\eqref{alphaXXZ} is proportional to the $T$-derivative of the magnetization along $x$ times the magnetic field and thus only contributes for $H_b \neq 0$. 

Importantly, the pressure dependence of the transverse and longitudinal coupling of the XXZ chain \eqref{SpinChain}, $J_\perp$ and $J_z$, respectively, are related to the ones of the Heisenberg exchange, $J_\mathrm{H}$, and the single-ion anisotropy, $D$, via Eqs.~\eqref{constants}. Thus, the thermal expansion at $H_b=0$ is determined by the pressure dependences of $\Theta_{\rm D}$, $D$ and $J_\mathrm{H}$. Again, the strength of the Schottky anomaly basically fixes $\partial D/\partial p_b$, and the low-energy peak then essentially determines the remaining parameter $\partial J_\mathrm{H}/\partial p_b$ according to Eq.~\eqref{alphaXXZ}. From a simultaneous best fit, shown as the red solid line in Fig.~\ref{fig2}(b), we obtain $\partial \ln \Theta_{\rm D}/\partial p_b = 0.01/{\rm GPa}$, 
\begin{align} \label{PressureDependence}
\frac{\partial \ln D}{\partial p_b} = -0.63/{\rm GPa},\mbox{ and }
\frac{\partial \ln J_\mathrm{H}}{\partial p_b} = 0.77/{\rm GPa}.
\end{align}
As is clearly seen in Fig.~\ref{fig2}(b), $\alpha_{\rm ph}$ only yields a small phononic background in this low-temperature range or, in other words, the measured thermal expansion is almost entirely of magnetic origin. From $J_\mathrm{H}/D \ll 1$, one might be tempted to expect that both bond correlators, Eqs.~\eqref{transCorr} and \eqref{longCorr}, contribute with a similar weight to $\alpha_{\rm XXZ}$. However, the pronounced minimum of $\alpha$ around $6$~K results in a relatively large value of $\partial D/\partial p_b$, which substantially influences the pressure dependence of $J_z$ and reduces the weight of the longitudinal $\langle S^z S^z \rangle$ bond correlator considerably. 
Whereas according to Eqs.~\eqref{constants}  the pressure dependence of $J_\perp$ is directly related to the one of $J_\mathrm{H}$, the pressure derivative of the longitudinal coupling,
\begin{align}
\frac{\partial J_z}{\partial p_b} =  \left(1- \frac{39}{4} \frac{J_\mathrm{H}}{D} \right)
\frac{\partial J_\mathrm{H}}{\partial p_b} + \frac{39}{8} \left(\frac{J_\mathrm{H}}{D} \right)^2 \frac{\partial D}{\partial p_b},
\end{align}
is sensitive to both quantities, $J_\mathrm{H}$ and $D$.
We find that the relatively large $\partial D/\partial p_b$ 
basically compensates the small factor $(J_\mathrm{H}/D)^2 \approx 0.01$, reduces the pressure dependence of $J_z$ and even drives it negative, 
$\partial J_z/\partial p_b \approx -0.1 \, \partial J_\perp/\partial p_b$. The longitudinal correlator thus barely contributes to $\alpha_{\rm XXZ}$. Moreover, our data reveal that $J_z/J_\perp$ would considerably decrease under uniaxial pressure $p_b$, such that Cs$_2$CoCl$_4$ would approach the XY limit even further. Having established the properties at zero field, we now turn to the magnetic field dependence and the discussion of the field-induced Ising transition. 
For the magnetic field dependence we need to determine the $g$ factor of the Hamiltonian \eqref{SpinChain}, which is fixed by the value of the critical field $H_b^{\rm cr}$ \cite{Caux2003}. The latter can be extracted from the positions of peaks in $c/T$ and $\alpha/T$, which obey a characteristic scaling $T \sim |H_b-H_b^{\rm cr}|^{\nu z}$ with $\nu z = 1$ as expected for Ising criticality that yields $H_b^{\rm cr} \approx 2$ T, see Fig.~\ref{fig1}. 
The existence of such peaks close to criticality is expected from very general considerations \cite{Zhu2003,Garst2005,Weickert2010}.

\begin{figure}
 \centering
 \includegraphics[width=1.\columnwidth]{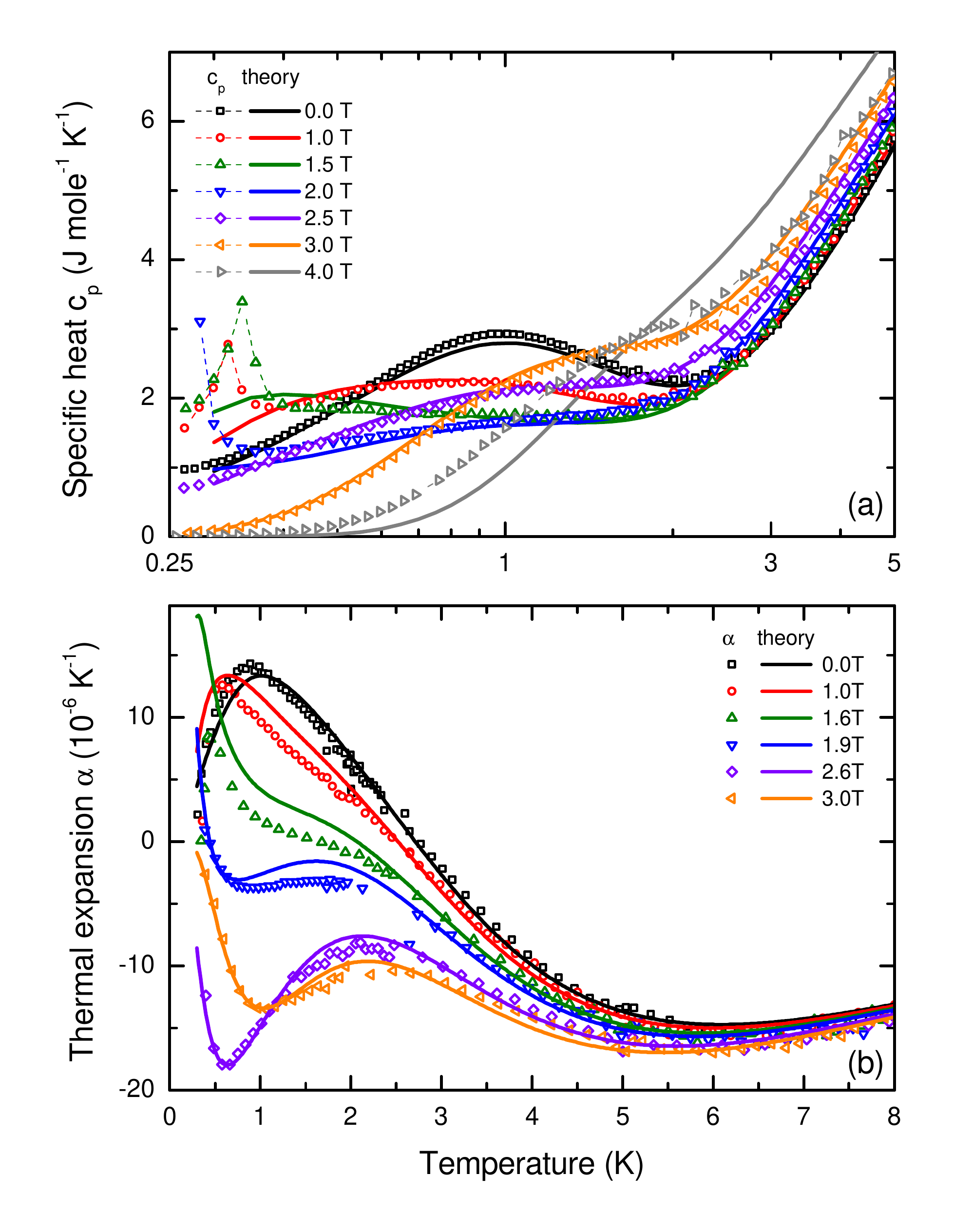}
 \caption{(color online) Specific heat and \textit{b}-axis thermal expansion of Cs$_2$CoCl$_4$ (symbols) for different magnetic fields $H_b$. The solid lines are best fits to theory, see text.}
 \label{fig3}
\end{figure}

As the extrapolation of the peak scaling suffers, however, from a relatively large error, we find it more convenient to determine  $g$ from the full set of specific heat curves at various fields, see Fig.~\ref{fig3}(a). From the best fit we obtain $g = 3.27$, which nicely agrees with the available magnetic moment $g \mu_\mathrm{B}/2 \approx 1.6 \mu_\mathrm{B}$ as inferred from neutron diffraction \cite{Kenzelmann2002}. Inverting the relation for the $g$ factors in Eqs.~\eqref{constants} this implies $g_{\frac{3}{2}} = 1.94$ given in Eqs.~\eqref{parameters}. As already emphasized in Ref.~\cite{Kenzelmann2002}, these values are, however, significantly smaller than expected from a  Curie-Weiss fit of the high-temperature susceptibility which yields $g_{\frac{3}{2}} \approx 2.4$~\cite{Figgis1964,McElearney1977,smit1979field}.

At $H_b \neq 0$, there is an additional contribution to $\alpha_{\rm XXZ}$ proportional to the pressure dependence of the effective $g$ factor, see Eq.~\eqref{alphaXXZ}. This term appears to be negligible, however, as the fits to the thermal expansion data, shown by the solid lines in Fig.~\ref{fig3}(b), result in $\partial g/\partial p_b\approx 0$.
From Eq.~\eqref{constants} one can relate $\partial g/\partial p_b$ to the pressure dependencies obtained in zero field,
\begin{align}\label{dgdp}
\frac{\partial g}{\partial p_b} = \left(2-\frac{3J_\mathrm{H}}{D}\right)\frac{\partial g_{\frac{3}{2}}}{\partial p_b}- 3 g_{\frac{3}{2}} \left(\frac{1}{D}\frac{\partial J_\mathrm{H}}{\partial p_b} - \frac{J_\mathrm{H}}{D^2}\frac{\partial D}{\partial p_b} \right), 
\end{align}
which allows to determine $\partial \ln g_\frac{3}{2}/\partial p_b\approx 0.26/{\rm GPa}$.

A striking observation is the sign change of the thermal expansion close to the critical field. 
Here, the singular part of $\alpha$ will be proportional to the change in entropy with respect to $H_b$, $\alpha_{\rm sing} \sim \partial S/\partial H_b$. The sign change of $\alpha$ seen in Fig.~\ref{fig3}(b) is related to the accumulation of entropy near the critical field and is, in fact, characteristic for quantum criticality \cite{Garst2005,Lorenz2008,Lorenz2007}.

In general, we find a good description of the experimental data in a large temperature and field range although there are deviations between theory and experiment visible, in particular, for the thermal expansion at finite field. The description in terms of a one-dimensional XXZ chain breaks down at the lowest temperature where the long-range order sets in and at higher fields when the two doublets start to become strongly entangled by the field, see Fig.~\ref{fig1}. This is most clearly illustrated in Fig.~\ref{fig3}(a) where the curve at $4$~T fails to provide a good description of the experimental data. Obviously, a finite entanglement is already present at lower fields and as a consequence, the field-induced splitting of the ground-state doublet at finite intermediate fields is less steep than in the limit $H_b \rightarrow 0$~\cite{smit1979field}. This may, at least partially, explain the significant deviation between the values of $g_\frac{3}{2}$ determined here and the corresponding high-temperature result.

In summary, specific heat and thermal expansion measurements on Cs$_2$CoCl$_4$ were presented for various magnetic fields along the $b$ axis. 
We demonstrated that they are well explained in terms of an effective spin-$\frac{1}{2}$ XXZ chain Hamiltonian in transverse magnetic field supplemented by a Schottky anomaly.
We were able to determine the corresponding coupling constants with high accuracy after exploiting that both contributions derive from the same parent Hamiltonian \eqref{Ham1}.
Whereas the Schottky anomaly is attributed to the thermal population of higher crystal field states, the virtual excitations of the same states result in important corrections to the spin chain which are essential for a consistent explanation of thermodynamics in Cs$_2$CoCl$_4$.

This work was supported by the Deutsche Forschungsgemeinschaft via SFB 608 and FOR 960.

\end{document}